\documentclass[journal,onecolumn]{IEEEtran}
\IEEEoverridecommandlockouts
\usepackage{cite}
\usepackage{amsmath,amssymb,amsfonts}
\usepackage{graphicx}
\usepackage{textcomp}
\usepackage{xcolor}
\usepackage{url}
\usepackage{booktabs}
\usepackage{array}
\usepackage{multirow}
\usepackage{comment}
\usepackage{float}

\begin{document}

\title{From Regulation to Engineering: Organizational Privacy Climate as Mediator}

\author{Oshrat Ayalon$^{1}$, Eran Toch$^{2}$, Irit Hadar$^{1}$, Michael Birnhack$^{3}$\\[0.6em]
\normalsize
$^{1}$Faculty of Computer and Information Science, University of Haifa, Israel\\
$^{2}$Faculty of Engineering, Tel Aviv University, Israel\\
$^{3}$Faculty of Law, Tel Aviv University, Israel
}

\maketitle

\begin{abstract}
Data protection legislation aims to protect privacy rights by regulating how information systems store and process personal information. While such legislation defines legal obligations, translating regulatory requirements into engineering practice is far from straightforward. As new frameworks such as the EU AI Act emerge, understanding these mechanisms becomes increasingly relevant. To investigate them, we conducted an online survey of 156 U.S. information system developers, examining the role of organizational privacy climate in the association between developers' perceptions of the GDPR and their privacy engineering practices. Using structural equation modeling, we found that organizational privacy climate mediates the relation between developers' perceived GDPR importance and their privacy attitudes, which in turn are associated with their privacy engineering practices. The mediation chain - regulation, climate, attitudes, practices - complements the direct effect of regulation on practice, revealing the organizational and attitudinal pathway through which regulatory intent reaches engineering behavior. We discuss implications for the next generation of privacy and AI regulation.
\end{abstract}


\section{Introduction}

Legislation efforts, such as the European Union's General Data Protection Regulation (GDPR) 
or the California Consumer Privacy Act (CCPA), aim to regulate the development of 
privacy-preserving systems. While such regulations define legal obligations, translating them 
into engineering practice is far from straightforward \cite{gurses2011}. Developers are 
responsible for embedding regulatory requirements into the systems they build, yet prior work consistently shows that familiarity with regulation alone does not automatically produce privacy-preserving practice \cite{balebako2014,alhazmi2021}. As privacy legislation expands 
across US states \cite{feld2020} and the EU AI Act enters force as the next major regulatory 
wave \cite{chun2024comparative}, understanding the \textit{mechanisms} through which 
regulation actually shapes engineering behavior becomes increasingly urgent. The GDPR, as a 
regulation that has been in force long enough for its effects to settle, offers an 
opportunity to study these mechanisms empirically -- and the lessons it offers are directly 
relevant to the regulatory landscape that follows.

To investigate these mechanisms, we look at the beliefs and decisions of software developers, as the values they embed in the design of information systems have a direct impact on how these systems manage privacy and personal data \cite{balebako2014,hadar2017}. Recent empirical work has identified key pieces of this puzzle: organizational culture and climate have been shown to shape privacy practices \cite{prybylo2024evaluating,iwaya2023privacy}, and even when developers are aware of regulatory requirements, translating them into compliant practice remains difficult \cite{horstmann2025sorry,klymenko2022understanding}. 
However, the \textit{mechanism} through which regulation reaches engineering practice -- and specifically the role of organizational factors as mediating forces -- has not been tested quantitatively. We address this gap by proposing and empirically testing a mediation model grounded in Social Cognitive Theory.

This study proposes a model for understanding the role of \textit{organizational privacy climate} -- the developers' perceived organizational expectations about privacy-related behavior -- in how the GDPR influences privacy engineering practices. To structure our model, we used Social Cognitive Theory (SCT) which suggests a bidirectional causation relationships structure between three determinants: (1) environmental factors, (2) personal factors; and (3) behavior \cite{bandura1986}. In our study context, the explored behavior is \textit{privacy engineering practices} as they refer to the developers' practices. We hypothesize that the former two determinants, environmental and personal factors, influence the behavior. To explore the GDPR effect on privacy engineering practices, our model explores the relationship between the \textit{GDPR's perceived importance}, as perceived by the developers, and privacy engineering practices. We hypothesize that the GDPR influences privacy engineering practices both directly and indirectly - through organizational privacy climate and developers' privacy engineering attitudes. Ultimately, we explore the influence on privacy engineering practices by the GDPR and organizational privacy climate (environmental factors), the developers' privacy engineering attitudes (personal factor), and the relationships between these environmental and personal factors.

We surveyed 156 information system developers to evaluate our model and used Structural Equation Modeling (SEM). All participants resided in the US and were recruited in the middle of 2020, approximately two years after the GDPR came into force. Our findings show that the developers' perceived GDPR's importance is directly and positively associated with the organizational privacy climate. The climate mediates the relation between the GDPR and the developers' privacy attitudes, which are associated with the developers' privacy engineering practices. Moreover, the developers' perceived GDPR's importance is directly and positively associated with the developers' privacy engineering practices. These results point to a deep influence of the GDPR on developers. Developers are not merely implementing privacy solutions to comply with the GDPR, but their privacy mindset is also influenced through their organization: developers that work in organizations with privacy-supportive organizational privacy climate are more likely to have privacy-supportive mindset.

\section{Background}

\subsection{Privacy Regulation: General Data Protection Regulation}

In 2018, the GDPR came into effect and changed data collection and processing in Europe and elsewhere. The main goals of the GDPR are to protect the rights and privacy of individuals in the EU and allow data flow through the EU to facilitate business \cite{schwerin2018}. One of the core aspects of the GDPR is Article 25: data protection by design and by default, which is based on the idea of Privacy by Design (PbD). The design principles and processes suggested by PbD aim to ensure that the organization considers privacy and data protection issues from the beginning of the design phase rather than after the fact \cite{cavoukian2009,langheinrich2001,drev2022}. The GDPR supports PbD with its transparency principle, thus providing the individuals with complete information and accountability, requiring the organizations to take responsibility for using personal data \cite{goddard2017}.

As anticipated \cite{li2019,safari2017}, the GDPR's impact extended well beyond the EU. Empirically, studies found changes in companies' privacy policies \cite{zaeem2020}, app permission practices \cite{momen2019}, and cookie implementations across US and non-EU websites \cite{sanchezrola2019}. At the legislative level, several US states enacted GDPR-inspired privacy laws \cite{feld2020}, reflecting the regulation's broader global reach \cite{mccabe2019}. Critically for the present study, these effects were not limited to EU-based organizations: US companies adjusted their systems and development processes in response to the GDPR, even without being legally bound by it. This extraterritorial influence makes US developers a particularly informative population for studying how regulatory intent travels beyond jurisdictional borders - they are affected by the regulation through organizational and market pressures rather than legal obligation, making the mechanisms of that influence more visible.

\subsection{Social Cognitive Theory and Developer Privacy Behavior}

In information systems, privacy possibilities and expectations are framed by the design of software architectures. Our study investigates how the law, organizational, and personal factors are associated with privacy engineering practices. Developers are responsible for translating business and legal requirements into concrete code and designing the software architectures that frame how these requirements are implemented. We use Social Cognitive Theory (SCT) to understand how the law influences privacy engineering practices, i.e., what forces are involved in the relationship between the law and privacy engineering practices. SCT suggests a bidirectional causation relationship between three determinants: cognitive and other personal factors, environmental influences, and behavior \cite{bandura1986}. The theory is widely used in Information Systems (IS) to model individual behavior \cite{compeau1999}, including developers' privacy perspectives \cite{hadar2017}. While SCT posits bidirectional relationships, reciprocal influences unfold over time \cite{bandura1986}. Therefore, we use SCT as a theoretical lens to structure our hypotheses in our cross-sectional design. In our model (summarized in Figure~\ref{fig:model}), we explore the influence on privacy engineering practices (behavior) of the GDPR's perceived importance and organizational privacy climate (environment), and the developers' privacy engineering attitudes (personal).

\subsection{Privacy regulation: The developers' perspective}

Former studies explored different aspects related to the implementation of privacy regulations by developers and closely related stakeholders. For example, studies explored designers' related perceptions \cite{luger2015}, differences in implementations between countries \cite{zanker2021}, and suggested concrete schemes to assist implementations \cite{aljeraisy2021}. In this subsection, we review these and other studies and point to the remaining gaps.

Several studies explored how developers implemented the GDPR principles and addressed its challenges. Across qualitative and quantitative work, a consistent picture emerges: developers are increasingly aware of the GDPR, yet awareness alone does not translate into compliant practice. Alhazmi and Arachchilage \cite{alhazmi2021,alhazmi2020} attributed this gap to unfamiliarity with the regulation and a lack of satisfying implementation techniques, though their relatively small samples limit broader generalization. Larger studies confirm that awareness has grown considerably -- Zanker et al. \cite{zanker2021} 
surveyed employees across eight EU countries and found most were aware of the GDPR, and Iwaya et al. \cite{iwaya2023privacy} reported increasing post-GDPR awareness across 30 practitioners in nine countries -- but they also confirm that translation to practice remains difficult. Iwaya et al. found that privacy engineering practices remain largely unstructured. Looking beyond regulation alone, Prybylo et al. \cite{prybylo2024evaluating} surveyed 362 team members across 23 countries and found that organizational culture, rather than regional regulation, primarily determines privacy practices -- pointing to organizational factors as a critical piece of the picture.

Contextualizing these results within our framework, we find a gap in understanding how the GDPR influences the system's design. While previous studies illuminate obstacles that hinder the GDPR's implementation, the broader factors shaping developers' privacy practices remain underexplored. Earlier work has examined personal \cite{tahaei2021} and organizational \cite{hadar2017,ayalon2017} factors that influence developers' privacy practices in general. Recent studies further suggest that translation from regulation to practice is mediated by organizational mechanisms. Klymenko et al. \cite{klymenko2022understanding} identified 
``Go-Betweens'' -- hybrid roles bridging technical and legal domains -- as critical for operationalizing vague regulatory requirements. Building on prior work, the current study examines how organizational and personal factors together shape the relationship between the GDPR and developers' privacy practices.

Introducing our first hypothesis, we explore the direct relationship between the GDPR and privacy engineering practices. In order to evaluate the GDPR from the developers' perspective, we measure the developers' perception of the GDPR's importance. Accordingly, we introduce our first hypothesis:

\noindent\textit{\textbf{Hypothesis 1}: GDPR's perceived importance is positively associated with privacy engineering practices.}

\subsection{Organizational Privacy Climate}

To further explore the hypothesized complex relationship between the GDPR and privacy engineering practices, we turn to the organizational influence. Previous work pointed to the impact of the organizational environment on the developers' privacy and security perceptions and practices \cite{arizon2021a,arizon2021b,ayalon2017,hadar2017}. Recent studies have further confirmed organizational privacy climate as a powerful force influencing practitioner behavior \cite{iwaya2023privacy,prybylo2024evaluating}. We built on what Hadar et al. \cite{hadar2017} defined as \textit{organizational privacy climate}, the organizational influence on developers' privacy perceptions and behaviors. In general, organizational climate involves employees' perceptions of the properties of the work environment, including the expectations of the organization regarding its employees' behavior as communicated via means such as policies, practices, routines, discourse, and rewards \cite{griffin2000,schneider2017,zohar2005}. In the context of safety climate, previous work pointed to safety rules as one of the components of the organizational climate \cite{mosly2019}. In our work, however, we are interested in understanding how the GDPR influences privacy engineering practices directly and indirectly via the organizational environment. Therefore, we explore its influence on the organizational privacy climate, not as one of the climate's constructs. Since the organizational climate was present before the GDPR appearance, we are interested in understanding how a new environmental factor -- privacy legislation -- is associated with the organizational privacy climate.

Accordingly, we propose the following  hypotheses:

\noindent\textit{\textbf{Hypothesis 2}: GDPR's perceived importance is positively associated with organizational privacy climate.}

\noindent\textit{\textbf{Hypothesis 3}: Organizational privacy climate is positively associated with privacy engineering practices.}

Beyond its direct influence on practices, organizational climate may also shape how 
developers think about privacy. Qualitative work has shown that the organizational 
environment influences developers' cognitive factors and personal attitudes toward 
privacy \cite{hadar2017,arizon2021b,iwaya2023privacy}. We therefore hypothesize:

\noindent\textit{\textbf{Hypothesis 4a}: Organizational privacy climate is positively 
associated with pro-privacy engineering attitudes.}

\subsection{Developers' Privacy Engineering Practices}

A growing body of work has investigated how developers approach privacy and what shapes their privacy-related decisions, attitudes, and practices.
Developers consistently report that they did not acquire formal privacy training and are unfamiliar with privacy tools, a finding that has persisted from early mobile-application research \cite{balebako2014} through more recent studies of practitioners across diverse contexts \cite{iwaya2023privacy,prybylo2024evaluating}. A key reason practices remain informal and under-implemented is the difficulty of translating high-level privacy requirements into concrete engineering work. Klymenko et al. interviewed 16 privacy professionals and identified ``Go-Betweens'' -- hybrid technical-legal roles -- as critical for translating vague regulatory requirements into actionable engineering tasks, finding that without them, developers and legal experts lack a shared understanding of what ``technical measures'' actually mean in practice \cite{klymenko2022understanding}. Similarly, Franke et al. surveyed open-source developers about their experience with the GDPR and identified the vagueness of regulatory requirements as one of the most reported challenges in practice \cite{franke2024exploratory}.

Beyond familiarity, developers' own privacy perceptions also shape their practices. Senarath and Arachchilage found in a lab study that developers did not use privacy best practices when they considered them unimportant \cite{senarath2018}. Sheth et al. explored users' and developers' privacy perceptions and found significant differences between them \cite{sheth2014}: developers were less concerned than users about data aggregation, and preferred technical mechanisms such as anonymization over privacy-law- or privacy-policy-based approaches. A persistent gap also remains between developers' privacy attitudes and their actual practices. Van den Linden et al. developed a scale to measure developers' attitudes towards their perceived personal data handling and found a clear mismatch \cite{linden2019}: developers were not bothered by collecting personal data from many users, yet at the same time perceived themselves as handling personal data correctly.

Researchers have explored ways to encourage developers to build privacy-preserving systems. Exploring the antecedents of PbD implementation, Bu et al. found that information systems engineers' PbD implementation is affected by social influence within their organizations \cite{bu2020,bu2021}. By ``social influence,'' Bu et al. referred to the engineers' perceived social pressure towards using PbD, finding that they are more likely to use PbD when they perceive that their professional surroundings prefer using PbD. Shilton used an 
ethnographic methodology in a computer science laboratory and characterized value levers: design practices that can encourage conversations and actions concerning ethical challenges \cite{shilton2013}. Yet individual-level interventions have shown limited effect: Horstmann et al. \cite{horstmann2025sorry} found in a lab study with 30 professional developers that even explicit privacy prompts and access to a privacy expert left the majority of GDPR implementations non-compliant.

Several studies have further identified organizational factors as central to developers' privacy behavior. Hadar et al. conducted a qualitative study exploring information systems architects' privacy decision-making processes. The study's findings indicate that three main factors influence software architects' privacy-related decisions and behaviors: (1) the architects' perceptions of privacy, (2) the organizational environment, and (3) the privacy-related technologies they use in practice \cite{hadar2017}. Following studies have further confirmed organizational privacy climate as a powerful force shaping developers' privacy behavior. Ayalon et al. surveyed developers and found that their privacy practices were directly influenced by the organizational privacy climate and by their perceptions of information control \cite{ayalon2017}. Tahaei et al. interviewed ``Privacy Champions,'' namely developers who promote users' privacy in their teams and organizations \cite{tahaei2021}, and found that Privacy Champions were motivated to encourage privacy by personal and organizational factors. Iwaya et al. conducted in-depth interviews with 30 practitioners across nine countries and identified organizational climate -- including leadership support and team norms -- as a central driver of privacy engineering behavior, with practices remaining largely informal in organizations lacking such climate \cite{iwaya2023privacy}. Most directly relevant to the present study, Prybylo et al. surveyed 362 software team members from 23 countries and explicitly compared the relative influence of organizational culture and regional regulation, finding that privacy practices are primarily shaped by organizational culture rather than by the regulatory regime under which developers operate \cite{prybylo2024evaluating}.

Integrating previous works on the topics of (a) privacy regulation, (b) organizational privacy climate, and (c) developers' privacy aspects, we propose the following hypotheses:

\noindent\textit{\textbf{Hypothesis 4b}: GDPR's perceived importance is positively associated with pro-privacy engineering attitudes.}

\noindent\textit{\textbf{Hypothesis 5}: Pro-privacy engineering attitudes are positively associated with privacy engineering practices.}

\begin{figure}[htbp]
\centering
\includegraphics[width=\columnwidth]{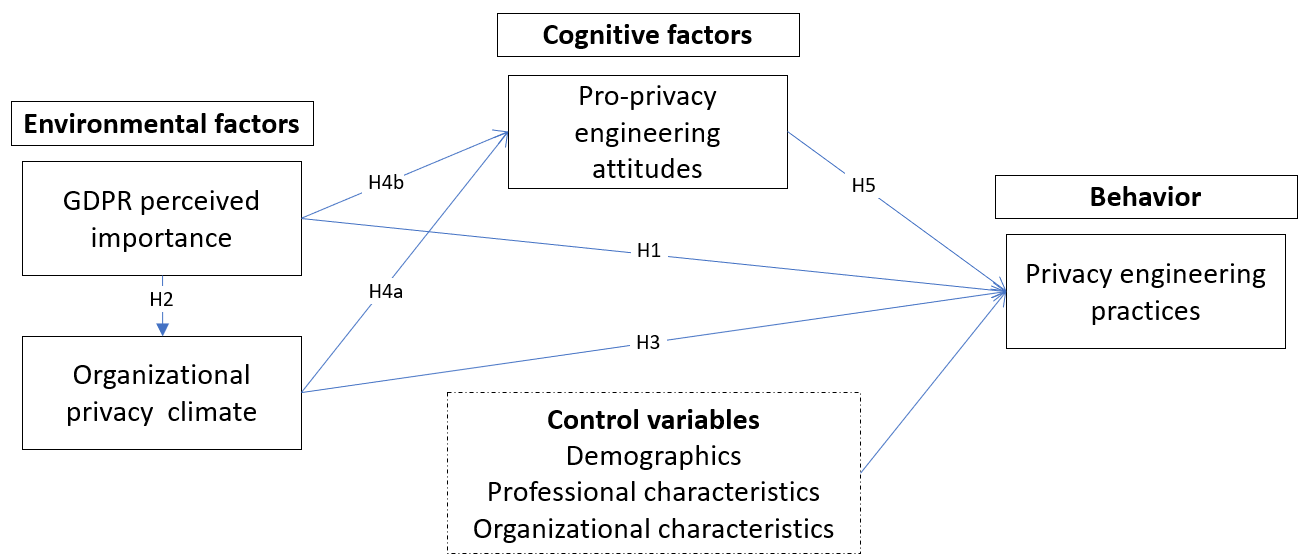}
\caption{Research model depicting the hypothesized relationships between GDPR perceived importance, organizational privacy climate, privacy engineering attitudes, and privacy engineering practices.}
\label{fig:model}
\end{figure}

Figure~\ref{fig:model} summarizes the study hypotheses and illustrates the hypothesized relationships among the research constructs.

\section{Methodology}

We conducted an online survey with 156 participants to reach the study's objectives. In this section, we describe our methodology and analysis approach.

\subsection{Measurements development}

We designed a survey to explore our research hypotheses. The measurement scales in the questionnaire were partially based on existing literature and partially developed by us. We used a 5-point Likert scale throughout the questionnaire, while the scale anchors differed between question sets. For example, in some cases, we asked about the developers' agreement with the provided statements, and in other cases, we asked about the frequency of used practices. The used measurement items are included in Appendix~A.

In this study, the examined behavior was measured by the developers' \textit{privacy engineering practices}. The questionnaire began by asking about the implementation of several privacy practices. The statements were developed mainly by us and were based on fair information practices (FIPs) \cite{gellman2016} and strategies based on PbD and on statements from Balebako et al.'s study \cite{balebako2014}. For example, in one statement, we asked about the frequency of ``Restricting access to personal data, so it is available only to the end-user.'' The answers ranged between ``No, I have never implemented it'' and ``Yes, I implement it often.''

We measured one cognitive-related variable: privacy engineering attitudes. To measure \textit{privacy engineering attitudes}, we developed the variable's statements based on Ayalon and Toch's Users' Perceived Systems' Privacy scale \cite{ayalon2019}. The statements in the Ayalon and Toch scale elicit the users' perceptions of a tested information system. In the current scale, we have made adaptations to ask the developers how they perceive a privacy-preserving system. For example, the developers were asked about their agreement with the following statement: ``Information systems should allow only authorized personnel to access users' personal information.''

To explore the impact of environmental factors on the developers' behavior, we measured two variables: organizational privacy climate and GDPR's perceived importance. We developed a scale to measure \textit{organizational privacy climate} by adapting Griffin and Neal's questionnaire \cite{griffin2000}. Griffin and Neal's questionnaire measures a workplace's safety climate, and we adjusted their safety questions to ask about privacy climate while considering the FIPs \cite{gellman2016}. For example, the developers were asked to indicate their agreement with the following statement: ``Management is sincere in its efforts to ensure end-users data privacy.'' Our second explored environmental variable was \textit{GDPR's perceived importance}. To measure the influence of GDPR, we asked about the developers' perceptions and attitudes toward the GDPR. They were asked about changes since the GDPR came into effect. For example, they were asked to indicate their agreement with the following statement: ``Since the GDPR, privacy issues are more frequently discussed at my work.''

In addition to cognitive and environmental variables, we controlled for demographic, professional-related, and organizational-related variables. To measure these variables we asked, for example, about the developer's hierarchical positions and years of experience, and organizational questions were about its size, headquarters location, and the continents where it operates.

The questionnaire also included additional scales: privacy knowledge, personal control and risk perceptions, privacy responsibility, and users' expectations. These variables are outside the scope of the current paper and are not included in the analysis. All items are listed in Appendix~A.

\subsection{Data collection}
To administer our survey, we used Qualtrics and CloudResearch\footnote{\url{https://www.cloudresearch.com/}}. CloudResearch is a platform that enables marketers and researchers to recruit specific respondents and is used in academic research \cite{mohamed2020}, similar to other online research panels used in developer privacy studies \cite{prybylo2024evaluating}. We used CloudResearch ``Managed research'' service to recruit our participants to ensure we reach our desired sample. We limited our selection to those who agreed with the following statement: ``I am currently working or have worked in the last three years or more, independently or as an employee, in the area of development of information systems or systems that manage information about people.'' We intended to recruit people who directly influence the system's development. Therefore, we asked our survey manager at CloudResearch to recruit participants with a background in Computer Science or Information Systems but to exclude individuals who work in Information Technology (IT).

The study was authorized by the institutional ethics review board (IRB) and was deployed between March and June 2020. Two hundred nine participants were qualified to participate: they were 18 years old or older and indicated that they were working in software development or information systems development. To avoid priming the participants, similar to the method used by Egelman and Peer \cite{egelman2015}, we did not mention ``privacy'' during recruitment to minimize selection bias. Based on the mean and standard deviation of the completion duration, we removed responses that took less than 3.3 minutes to complete, resulting in 185 responses. Next, we analyzed whether a straight-lining case exists per each set of questions, in which the participant answered the same answer for all the questions. We removed responses that included four or more cases of straight-lining, which resulted in 164 responses. Lastly, we excluded participants who were not US residents, resulting in a final sample of 156 participants. After the data removal, the mean completion duration was 11.9 minutes, with a median of 8.5 minutes.

\subsection{Data Analysis}

To evaluate our questionnaire validity, in light of our explored model, we used several commonly used heuristics in IS research \cite{straub2004}. 
We performed Exploratory Factor Analysis (EFA) using Promax rotation to evaluate constructs validity. We then performed Confirmatory Factor analysis (CFA) to evaluate the EFA results. Lastly, we evaluated the resulted constructs' internal consistency using Composite Reliability (CR) and their discriminant validity using Average Variance Extracted (AVE). After ensuring the constructs' reliability and validity, we used structural equation modeling (SEM) to evaluate our suggested model and hypotheses.

\subsubsection{Measurement development}

SEM is widely used in Information Systems (IS) research and enables the evaluation of the measurements and the paths' model into one unified model \cite{gefen2011}. With SEM, researchers can perform path analytic modeling with unobserved (latent) variables \cite{chin1998}.

Based on our theoretical model, we specified four factors in the EFA, corresponding to the four constructs of interest: GDPR perceived importance, organizational privacy climate, privacy engineering attitudes, and privacy engineering practices.
We followed Hair et al.'s suggestion \cite{hair1998} and used 0.45 as a cutoff to include questions, and removed one question as a result (``Privacy issues are frequently discussed during staff meetings''). The final EFA resulted in 22 questions, with the lowest loading value of 0.525, and four factors: developers' privacy engineering attitudes, legal influence, organizational privacy climate, and privacy engineering practices. The four factors predicted 52.9\% of the variance. See Appendix~B for the table presenting the loadings and cross-loadings of the final included measures and questions.

In the next analysis step, we took Anderson and Gerbing's two-step approach, which calls first to assess the measurement model using confirmatory factor analysis (CFA) before evaluating the suggested model using SEM \cite{anderson1988}. To assess the fit of the measurement model and the structural model, we used five indices recommended by Hu and Bentler \cite{hu1999}. The results of the model evaluation showed that our data supported the model, as all the values follows the recommended cut-offs \cite{awang2012,hooper2008,vandenberg1994}: RMSEA = 0.063, SRMR = 0.063, CFI = 0.921, TLI = 0.91.

Lastly, we evaluated the resulted constructs' internal consistency using composite reliability (CR) analysis. All values were higher than 0.7, which is the suggested cutoff value \cite{hair2016}. We measured the constructs' discriminants validity using average variance extracted (AVE). In all cases, the square root of the variable is greater than the latent variables, pointing to adequate discriminant validity \cite{fornell1981}. 
See Appendix~C, Table~\ref{tab:cr} for the constructs' internal consistency and discriminant validity.

\section{Results}

\subsection{Sample and Main Variables' Descriptive Analysis}

Our final sample consisted of 156 participants. All of them were residents of the US, and most were aged 25-44 (59\%). The sample comprised 55\% women, and 44\% men. The sample's detailed demographics and professional positions are represented in Table~\ref{tab:demo}. 
Considering organizational characteristics, almost all participants reported that the headquarter of the organization they are working at is based in the US (N = 150). 
To learn where the organization operates, we provided a list of continents/areas 
and asked the participants to check all the continents/areas where the organization operates. To analyze the data, we considered answers that included ``Europe,'' possibly along with other areas, as ``Europe''; answers that included only ``North America'' as ``North America''; and other answers (N = 17) as ``Other.'' Accordingly, about sixty percent of the organizations operate in the US only (N = 94), 28.9\% of the organizations operate in Europe (N = 45), and the rest of the organizations operate globally (N= 17).

\begin{figure*}[htbp]
\centering
\includegraphics[width=\textwidth]{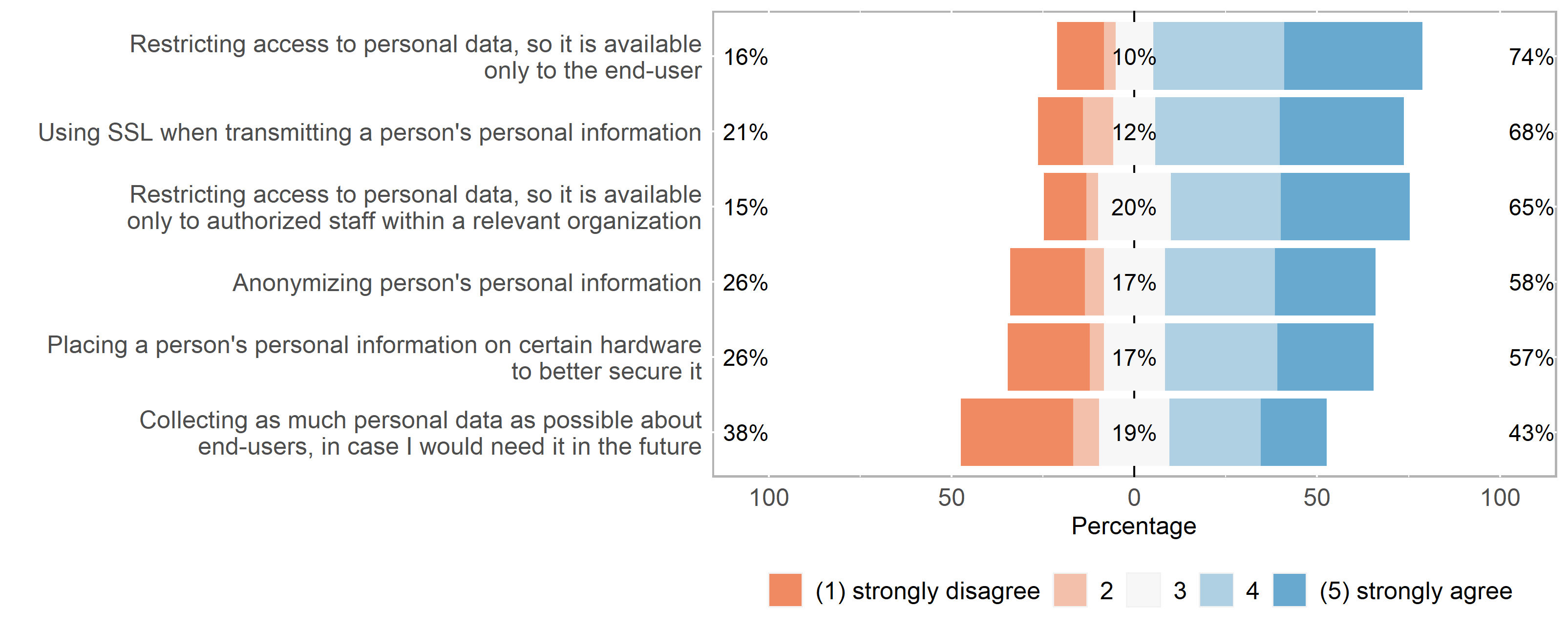}
\caption{Descriptive results of privacy engineering practices.}
\label{fig:practices}
\end{figure*}

The most used \textit{privacy engineering practice}, with 74\% of the participants answering they implement this feature occasionally or often (4 or 5 on Likert scale, respectively), was ``Restricting access to personal data, so it is available only to the end-user.'' Except for one feature, 57\% or more of the participants reported using all the features occasionally or often. The only feature that did not follow this trend was: ``Collecting as much personal data as possible about end-users, in case I would need it in the future.'' In this case, 43\% of the participants reported implementing this feature, and 38\% reported that they were \textit{not} implementing it. See Figure~\ref{fig:practices}, which demonstrates the distribution of the participants' responses.

In the questions that explored \textit{privacy engineering attitudes}, we saw a general agreement between the developers. Most developers agreed with the statement that ``Information systems should allow only authorized personnel to access users' personal information.'' 87\% of the participants agreed or strongly agreed (4 or 5 on the Likert scale, respectively) with this statement. The statement with which fewest participants agreed upon, and yet was highly agreed, was ``Information systems should allow users to control which information about them is being collected.'' 79\% of the participants either agreed or strongly agreed with this statement.

The findings regarding the \textit{organizational privacy climate} show that the participants generally perceive the privacy climate in their organization as positive. Most developers agreed with this statement: ``Management is sincere in its efforts to make sure personal data is used only for the purpose that was approved by the end-user.'' 79\% of the participants agreed or strongly agreed with the statement. The two statements upon which fewest participants agreed were ``The employees are equipped with enough privacy solutions for privacy issues (e.g., anonymization techniques)'' and ``Privacy issues are frequently discussed during staff meetings.'' 67\% and 65\% of participants agreed or strongly agreed with the statements, respectively.

The \textit{GDPR's perceived importance} results pointed to a general agreement with the statements. For example, most participants agreed or strongly agreed with this statement: ``Since the GDPR, I feel that people care more about their privacy.'' 75\% of the participants either agreed or strongly agreed with the statement. The statement with which fewest participants agreed or strongly agreed was: ``Since the GDPR, we are collecting less personal information.'' 55\% of the participants agreed or strongly agreed with the statement. In this set of questions, we also asked about the participants' familiarity with the CCPA, as they were US residents. As shown in Appendix~D, the results show that participants tended to be more 
familiar with the GDPR (65\% of the participants either agreed or strongly agreed that they are familiar with the GDPR) than with the CCPA (60\%). Our results seem to contradict Alhazmi and 
Arachcilage \cite{alhazmi2021} results, which pointed to the non-EU developers' unfamiliarity with the GDPR. However, their study did not mention where the non-EU developers were from, 
making it hard to compare to our US sample.

\begin{table}[htbp]
\caption{Sample's demographics}
\label{tab:demo}
\centering
\begin{tabular}{|l|l|r|}
\hline
\textbf{Variable} & \textbf{Value} & \textbf{\%} \\
\hline
\multirow{6}{*}{Age} & 18--24 & 12.2 \\
 & 25--34 & 29.5 \\
 & 35--44 & 29.5 \\
 & 45--54 & 17.3 \\
 & 55--64 & 9.6 \\
 & 65+ & 1.9 \\
\hline
\multirow{3}{*}{Gender} & Women & 55.1 \\
 & Men & 44.2 \\
 & Prefer not to answer & 0.6 \\
\hline
\multirow{7}{*}{Position} & Member of dev. team & 30.1 \\
 & Head of dev. team & 28.9 \\
 & Head of dev. organization & 10.9 \\
 & Works mostly by self & 10.3 \\
 & Requirements specification team & 9.0 \\
 & Design team & 8.3 \\
 & Student / Other & 2.5 \\
\hline
\multirow{3}{*}{HQ location} & United States & 96.2 \\
 & Other & 3.8 \\
\hline
\multirow{3}{*}{Org. operates in} & North America & 60.3 \\
 & Europe & 28.9 \\
 & Other & 10.9 \\
\hline
\multirow{5}{*}{Org. size} & 1 & 3.2 \\
 & 2--100 & 20.5 \\
 & 101--1{,}000 & 36.5 \\
 & 1{,}000--10{,}000 & 28.9 \\
 & More than 10{,}000 & 10.9 \\
\hline
\end{tabular}
\end{table}

\subsection{Structural Equation Modeling}

\begin{figure}[htbp]
\centering
\includegraphics[width=\columnwidth]{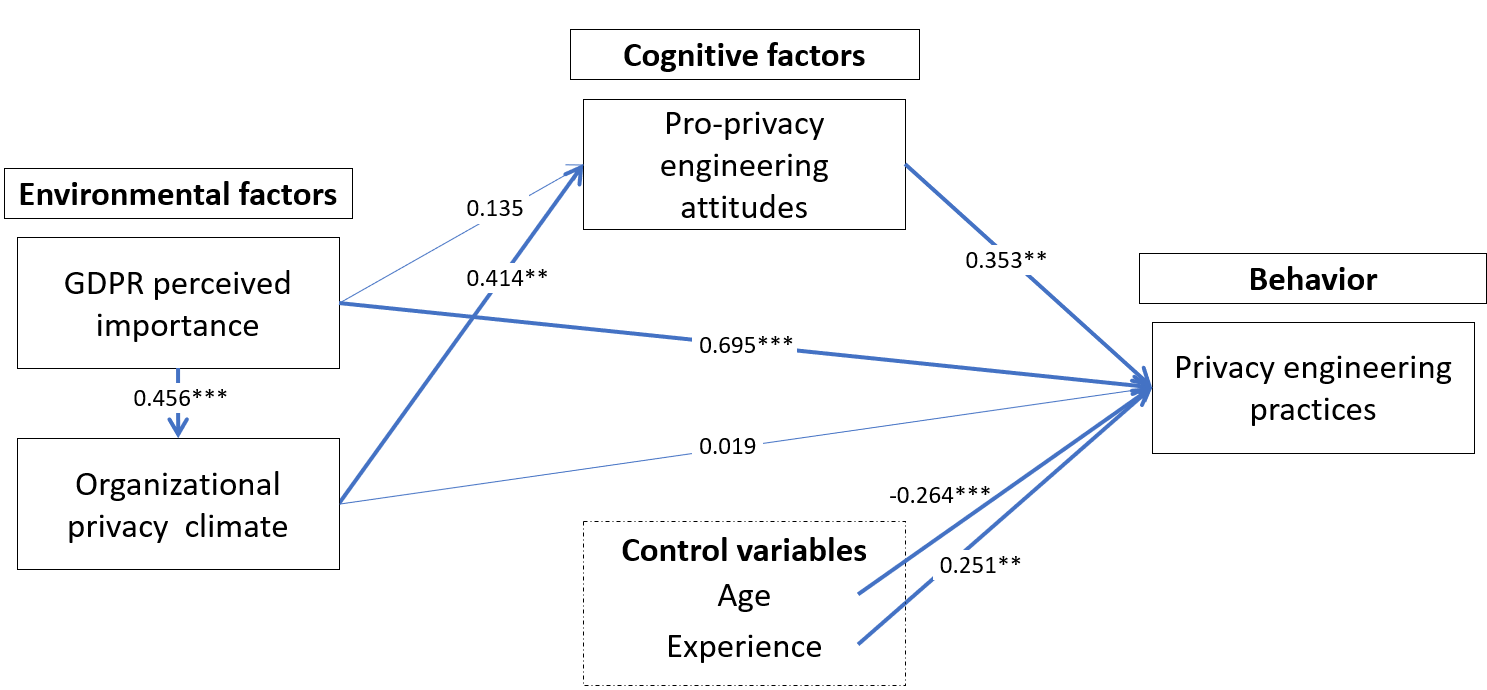}
\caption{Resulted model. Note: Standardized coefficients. * $p < 0.05$, ** $p < 0.01$, *** $p < 0.001$.}
\label{fig:sem}
\end{figure}

We used SEM to evaluate the hypothesized model, as depicted in Figure~\ref{fig:model}. During model development, two practice items were removed to improve model fit: the fifth item due to high uniqueness (u=0.71), and the sixth item as it represents an anti-privacy behavior conceptually distinct from the pro-privacy practices scale. 

Next, we evaluated a model that included all the control variables. However, the resulting model did not have an acceptable fit. Therefore, we removed non-significant control variables \cite{jia2015}: developers' hierarchical position, education, gender, organization's size, organization's headquarters' location, and the continents where the organization operates. The final model had an acceptable fit \cite{awang2012,hooper2008,vandenberg1994}: RMSEA = 0.054, SRMR = 0.067, CFI = 0.924, TLI = 0.916.

The results show a direct positive association between \textit{GDPR's perceived importance} and \textit{privacy engineering practices} ($\beta = 0.695$, $p < 0.001$), supporting H1. As the GDPR was considered more influential, the developers used more privacy practices in their work, with this association being the strongest among the model's paths. In addition to the observed direct effect, the results showed an indirect effect of the GDPR perceived importance through the \textit{organizational privacy climate}. Perceiving the GDPR as influential was associated with a more privacy-preserving organizational privacy climate ($\beta = 0.456$, $p < 0.001$), supporting H2. However, exploring the association between \textit{GDPR perceived importance} and \textit{privacy engineering attitudes}, we did not find a significant effect ($\beta = 0.135$, $p = 0.225$), rejecting H4b.

Examining the \textit{organizational privacy climate}, we did not find a significant direct association between organizational climate and \textit{privacy engineering practices} ($\beta = 0.019$, $p = 0.891$), rejecting H3. However, we did observe an indirect positive effect through the developers' \textit{privacy engineering attitudes}. As the organizational climate was more privacy-preserving, the developers' attitudes towards information systems design were also more privacy-preserving ($\beta = 0.414$, $p = 0.002$), supporting H4a. 
To formally test this indirect effect, we used bootstrapping with 1,000 samples \cite{preacher2008}. The indirect effect of GDPR perceived importance on privacy engineering practices through organizational privacy climate and privacy engineering attitudes was significant (indirect effect = 0.067, SE = 0.038, 95\% CI [0.023, 0.233]), supporting the proposed mediation chain.
Lastly, the results show a significant direct positive association between the developers' \textit{privacy engineering attitudes} and their \textit{privacy engineering practices} ($\beta = 0.353$, $p = 0.001$), supporting H5.

Exploring the effect of control variables on privacy engineering practices, we found that two variables had a significant effect: the developer's age and experience. Younger and more experienced developers were more likely to use more privacy practices, similar to Bu et al. \cite{bu2020} results (age: $\beta = -0.264$, $p < 0.001$; experience: $\beta = 0.251$, $p = 0.002$).

\section{Discussion}

This work aimed to investigate how the GDPR shapes US developers' privacy engineering practices, and through which mechanisms. To this end, we asked whether and how the developers' GDPR perceived importance is associated with their privacy practices. We hypothesized that the association involved a complex structure of relationships between environmental and personal factors. Our findings point to the influence of the GDPR beyond its binding jurisdiction, consistent with prior work showing that regulation shapes systems design well beyond its legal borders \cite{momen2019,zaeem2020,sanchezrola2019}. Where prior work evaluated specific system-level measures -- for example, Sanchez-Rola et al. examined websites' tracking implementation and found that US websites were impacted by the GDPR similarly to European ones \cite{sanchezrola2019} -- our study was interested in exploring the GDPR's impact \textit{on} the organization and the mechanisms that drive its effect. To examine the effect of privacy legislation on how organizations design information systems, researchers have either investigated the management level \cite{bamberger2011} or, as in our and other studies, focused on developers throughout the organization 
\cite{alhazmi2021,alhazmi2020,ayalon2017,iwaya2023privacy,prybylo2024evaluating,horstmann2025sorry,klymenko2022understanding,franke2024exploratory}. 
Prior work has identified organizational factors and personal attitudes as influences on developer privacy behavior, yet the quantitative relationships between regulation, organizational climate, personal attitudes, and engineering practices have not been jointly tested.

We found that the extent to which developers perceived the GDPR as meaningful is associated with their privacy practices. Our findings show both direct and indirect effects. Beyond the direct association between the GDPR's perceived importance and privacy practices, the results point to a deeper influence structure. We observed an indirect association between the GDPR perceived importance and the developers' privacy engineering attitudes, with organizational privacy climate as a mediator. We found that the developers were not merely directly affected by the GDPR, i.e., by following the regulations per se, but that their privacy mindset was also indirectly affected by the organizational climate in which they work.

\textbf{GDPR perceptions.} Prior work has documented that the GDPR is costly and difficult to implement, yet has real behavioral effects on developers and organizations. Studies found that while developers and organizations struggle with compliance, the regulation has raised awareness about personal data handling and prompted concrete changes in engineering practice \cite{mangini2020,franke2024exploratory}. Iwaya et al. 
\cite{iwaya2023privacy} similarly found that practitioners report the regulation as a significant external pressure that has shifted organizational and engineering behavior post-GDPR. Our results contribute to this picture by quantitatively measuring developers' perceived importance of the GDPR, and finding that when developers consider the regulation meaningful, the systems they design are more privacy-preserving. The perceived importance of regulation is not only an attitudinal outcome -- it is an antecedent of engineering behavior.

\textbf{Organizational privacy climate.} We found that organizational privacy climate mediated the relation between the developers' GDPR perceived importance and their privacy engineering attitudes, which were in turn directly associated with privacy engineering practices. Prior qualitative work consistently points to organizational climate as a powerful shaping force on how developers think and act on privacy, with top management commitment and clear organizational signals identified as particularly critical \cite{arizon2021b,iwaya2023privacy}. Our findings support this direction quantitatively, while also specifying that climate shapes developer attitudes rather than acting directly on practices. This indirect pattern also resonates with findings from the broader security compliance literature, where organizational factors have similarly been found to shape employee behavior through cognitive mediators rather than directly \cite{hong2022}. At the organizational level, dedicated roles bridging technical and legal domains may be one concrete way through which a supportive privacy climate is built and maintained \cite{klymenko2022understanding}. Somewhat consistent with our null finding on 
the direct climate--practices path, Prybylo et al. \cite{prybylo2024evaluating} found that the presence of a Chief Privacy Officer correlated with developer confidence in their organization's privacy measures but not with privacy outcomes -- though their measures and models differ considerably from ours, this pattern broadly resonates with our finding that organizational climate does not directly produce privacy engineering practices.

This picture points to developers' privacy attitudes as the critical link between the organizational environment and actual engineering behavior. While regulation reaches practice both directly and through organizational climate, it does not directly shape developers' professional attitudes -- that path runs through organizational climate first. Iwaya et al. \cite{iwaya2023privacy} observed a positive shift in practitioners' privacy mindset, attributing it in part to the growing pressure of privacy regulations to adjust 
engineering practices. Our results give this observation quantitative form, showing that the path from GDPR perceived importance to attitudes runs through climate. Strengthening the organizational privacy climate therefore emerges as a concrete leverage point -- connecting regulatory intent to developers' professional mindset, and through it, to engineering practice.

Our null finding on the direct climate-practices path aligns with Hong and Furnell \cite{hong2022} but contradicts prior work on privacy and security climate \cite{arizon2021a,ayalon2017}, which found significant direct associations between organizational climate and practices. Two factors may explain this difference. First, Arizon-Peretz et al. \cite{arizon2021a} focused on organizational \textit{security} climate rather than privacy climate. While these are closely related, differences in findings are to be expected. Second, while Ayalon et al. \cite{ayalon2017} studied organizational privacy climate, they examined a model which did not include a legal aspect. Since we observed a strong association between the legal aspect and privacy practices in our study, it may diminish the direct effect of organizational climate -- the climate's influence in our model operates indirectly, through developers' attitudes, rather than directly on practices.

\textbf{Lessons for the AI Act and the next generation of regulation.} Although our data were collected in the context of the GDPR, the mechanism we identify is not GDPR-specific. As the EU AI Act enters force \cite{chun2024comparative} and comparable AI and privacy laws expand across US states, organizations and regulators face a structurally similar challenge: moving regulation off the page and into engineering practice. Prior work and our own findings agree that regulation works -- it puts pressure on organizations, changes 
development processes, and has a direct effect on practice \cite{mangini2020,franke2024exploratory}. But our findings also show that this direct effect is not the whole story. When developers perceive regulation as meaningful, the organizational climate amplifies that signal -- shaping professional attitudes and, through them, making privacy-preserving practice more deeply embedded rather than merely reactive.Empirical work on AI ethics points to the same gap in the making -- ethical risk assessment tends to be a done-once activity rather than a continuous process, with principles remaining too abstract to translate directly into engineering decisions \cite{lu2022software}. For the AI Act, this points to investing in an organizational AI ethics climate analogous to the privacy climate we studied. Treating the AI Act purely as a legal compliance exercise risks reproducing the gap our findings reveal -- a regulation that exists, and a workforce whose mindset has not moved. 

\subsection{Limitations and Future Work}

This study has several limitations. Our sample is limited to 156 US-based developers, which limits generalizability to other regions and cultural contexts. Regarding GDPR relevance, approximately 30 percent of the participants worked in organizations operating in Europe and thus directly subject to the regulation, and 65\% reported familiarity with GDPR principles, suggesting meaningful exposure across the sample. Replication studies in other settings would strengthen the generalizability of the proposed mediation chain. 
In addition, the sample is skewed toward women relative to the broader developer population, though we did not find gender effects in our analysis. 

Participants were recruited through CloudResearch's Managed Research service; while crowdsourcing platforms have been used in software engineering research \cite{prybylo2024evaluating}, standard concerns about self-selection apply, where developers interested in privacy may have self-selected into this study at higher rates, potentially inflating our measures. This was mitigated through managed recruitment, the deliberate avoidance of privacy-related language during recruitment \cite{egelman2015}, attention checks, and screening for direct experience with personal data systems.

Data were collected in 2020, approximately two years after GDPR enforcement, reflecting a particular moment in the regulation's history. However, the mediation mechanism we identify - organizational climate shaping attitudes, which in turn shape practices - is a structural organizational and psychological process unlikely to reverse over time. Moreover, recent work suggests that developer privacy awareness has continued to grow in the post-GDPR period \cite{iwaya2023privacy}, suggesting that if anything, the perceived importance of regulation may be stronger today than at the time of our study. Finally, the cross-sectional self-report design limits causal inference, and reported practices may differ from actual implementations \cite{linden2019,horstmann2025sorry}.

Two directions appear especially promising for future work. Replicating the model in other regulatory contexts -- particularly the EU AI Act, once organizations have had time to respond to it -- would test whether the same mediation chain operates for AI ethics climate as for privacy climate. Examining what specific organizational practices translate into stronger privacy climate would also move the discussion from establishing that climate 
matters to understanding how to build it.

\section{Conclusion} 

This study examined how the GDPR shapes developers' privacy engineering practices and through which mechanisms. Using structural equation modeling on data from 156 developers from the US, we found that the GDPR's perceived importance reaches engineering practice both directly and indirectly -- through an organizational privacy climate that shapes developers' professional attitudes. The mediation chain -- regulation, climate, attitudes, 
practices -- is the contribution we offer to the literature on developer privacy. It complements qualitative and descriptive findings on the role of organizational culture by specifying how regulation and organizational climate interact to produce privacy engineering behavior.

As privacy regulation matures and new frameworks such as the EU AI Act enter force, our results suggest that regulation has a real and direct effect on practice -- and that organizational privacy climate amplifies and deepens that effect by shaping developers' professional mindset. The two work together. Future work should examine what specific organizational conditions strengthen privacy climate, and whether the same mediation 
chain operates in the context of AI regulation -- a question our model was not designed to answer, but that the arrival of the AI Act makes increasingly relevant.

\noindent\textit{\textbf{AI disclosure:} The authors used AI writing assistance (Claude, Anthropic) 
in preparing this manuscript. All content was reviewed and verified by the authors, who 
take full responsibility for the accuracy and integrity of the work.}

\bibliographystyle{IEEEtran}

\bibliography{refs}

\onecolumn
\appendices

\section{Measurement items}
\label{app:items}

All items measured on a 5-point Likert-type scale.

\textbf{Privacy engineering practices}
\begin{itemize}
\item Restricting access to personal data, so it is available only to the end-user.
\item Restricting access to personal data, so it is available only to authorized staff within a relevant organization.
\item Anonymizing person's personal information.
\item Using SSL when transmitting a person's personal information.
\item Placing a person's personal information on certain hardware to better secure it.
\item Collecting as much personal data as possible about end-users, in case I would need it in the future.
\end{itemize}

\textbf{Privacy engineering attitudes}
\begin{itemize}
\item Information systems should make it clear to the users for which purposes their personal information is being collected.
\item Information systems should let users control which personal information is being shared with other companies.
\item Information systems should allow only authorized personnel to access users' personal information.
\item Information systems should allow users to control which information about them is being collected.
\item Information systems should be clear about the period in which the users' personal information is being retained.
\end{itemize}

\textbf{Privacy knowledge}
\begin{itemize}
\item Technical textbooks
\item Courses devoted to privacy during my academic studies
\item Courses that covered privacy during my academic studies
\item In-house corporate training in my current organization
\item External training (outside of my organization)
\item Web resources such as forum messages, online articles or blog posts
\item Privacy specialist or Chief Privacy Officer
\item Lawyers
\end{itemize}

\textbf{Personal control}

\begin{quote}
Before answering these items, participants were asked to think about several websites and apps they commonly use (e.g., Facebook, Instagram, Twitter, TikTok, Amazon, Google, Reddit, WhatsApp, Snapchat, Tinder) and their access to their personal information. ``These services'' in the items below refers to that set of apps and websites.
\end{quote}

\begin{itemize}
\item I think I have control over what personal information is released by these services.
\item I believe I have control over how personal information is used by these services.
\item I believe I have control over what personal information is collected by these services.
\item I believe I can control my personal information provided to these services.
\end{itemize}

\textbf{Personal risk}
\begin{itemize}
\item In general, it would be risky to give personal information to these services.
\item There would be a high potential for privacy loss associated with giving personal information to these services.
\item Personal information could be inappropriately used by these services.
\end{itemize}

\textbf{Organizational privacy climate}
\begin{itemize}
\item Management is sincere in its efforts to secure end users' data against malicious cyber-attacks from outside of the organization.
\item Management is sincere in its efforts to ensure end-users' data privacy.
\item In my workplace, I know exactly who I should turn to when handling a privacy-related issue.
\item Privacy issues are frequently discussed during staff meetings.
\item The employees are equipped with enough privacy solutions for privacy issues (e.g., anonymization techniques).
\item Management is sincere in its efforts to make sure personal data is used only for the purpose that was approved by the end-user.
\end{itemize}

\textbf{GDPR perceived importance}
\begin{itemize}
\item Since the GDPR, privacy issues are more frequently discussed at my work.
\item Since the GDPR, I feel that people care more about their privacy.
\item Since the GDPR, we are collecting less personal information.
\item Since the GDPR, we provide more information to users about their personal information.
\item Because of the GDPR, I learned new privacy techniques that I can apply in my work.
\item I am familiar with the GDPR principles.
\item I am familiar with the CCPA principles.
\end{itemize}

\textbf{Privacy responsibility}
\begin{itemize}
\item I often feel responsible for privacy in the systems I develop.
\item I would be held personally responsible if people would complain about privacy issues in the systems I develop.
\item I should think about possible privacy consequences while developing a system.
\item I should take care of people's privacy in the systems I develop.
\item The way I design systems can affect people's privacy.
\item I feel that I can make a difference regarding people's privacy in the systems I develop.
\item I feel that if I argue for better privacy in my organization, I will be heard.
\end{itemize}

\textbf{Users' expectations}
\begin{itemize}
\item I think that I can accurately predict whether users would consider a system as privacy-intrusive.
\item I think that I will have similar perceptions as the end-users of whether a system is privacy-intrusive.
\item I think that if I consider a system as privacy-intrusive, users would consider it as privacy-intrusive as well.
\item I think that if I consider a system as privacy-respectful, users would consider it as privacy-respectful as well.
\end{itemize}

\section{Loadings and cross-loadings of measures}
\label{app:loadings}

\begin{table}[htbp]
\caption{Loadings and cross-loadings of measures - EFA process}
\label{tab:loadings}
\centering
\begin{tabular}{lrrrr}
\toprule
 & \textbf{Practices} & \textbf{Attitudes} & \textbf{Climate} & \textbf{GDPR} \\
\midrule
practice1 &  \textbf{0.84} &  0.03 & -0.06 & -0.04 \\
practice2 &  \textbf{0.79} &  0.13 &  0.12 & -0.13 \\
practice3 &  \textbf{0.71} &  0.01 &  0.08 &  0.04 \\
practice4 &  \textbf{0.61} &  0.00 & -0.17 &  0.14 \\
practice5 &  \textbf{0.53} & -0.12 &  0.15 &  0.04 \\
practice6 &  \textbf{0.53} & -0.17 &  0.03 &  0.14 \\
\midrule
attitude1 & -0.08 &  \textbf{0.84} &  0.09 &  0.11 \\
attitude2 &  0.01 &  \textbf{0.76} & -0.12 &  0.10 \\
attitude3 &  0.08 &  \textbf{0.76} &  0.06 & -0.09 \\
attitude4 &  0.13 &  \textbf{0.71} & -0.11 & -0.06 \\
attitude5 &  0.01 &  \textbf{0.76} &  0.13 & -0.02 \\
\midrule
climate1  &  0.10 &  0.03 &  \textbf{0.77} & -0.07 \\
climate2  &  0.07 &  0.04 &  \textbf{0.81} & -0.16 \\
climate3  & -0.10 &  0.00 &  \textbf{0.55} &  0.13 \\
climate5  & -0.05 & -0.04 &  \textbf{0.62} &  0.15 \\
climate6  & -0.05 &  0.04 &  \textbf{0.73} &  0.05 \\
\midrule
gdpr1     & -0.04 &  0.01 &  0.02 &  \textbf{0.76} \\
gdpr2     & -0.04 &  0.07 &  0.10 &  \textbf{0.60} \\
gdpr3     & -0.10 & -0.09 & -0.07 &  \textbf{0.81} \\
gdpr4     &  0.16 & -0.04 & -0.01 &  \textbf{0.63} \\
gdpr5     &  0.00 &  0.09 & -0.07 &  \textbf{0.75} \\
gdpr6     &  0.08 &  0.10 & -0.01 &  \textbf{0.66} \\
\bottomrule
\end{tabular}
\end{table}

\section{Internal consistency and discriminant validity of constructs}
\label{app:cr}

\begin{table}[H]
\caption{Composite reliability (CR), average variance extracted (AVE), and discriminant validity. \\Internal consistency (square root of AVE) shown in bold on the diagonal.}
\label{tab:cr}
\centering
\begin{tabular}{lrrrrrr}
\toprule
 & CR & AVE & practices & attitudes & climate & gdpr \\
\midrule
practices & 0.83 & 0.56 & \textbf{0.75} & & & \\
attitudes & 0.89 & 0.61 & 0.43 & \textbf{0.78} & & \\
climate   & 0.83 & 0.50 & 0.40 & 0.41 & \textbf{0.71} & \\
GDPR      & 0.86 & 0.50 & 0.61 & 0.30 & 0.51 & \textbf{0.71} \\
\bottomrule
\end{tabular}
\end{table}

\section{Familiarity with privacy legislation}
\label{app:familiarity}

\begin{figure}[H]
\centering
\includegraphics[width=\columnwidth]{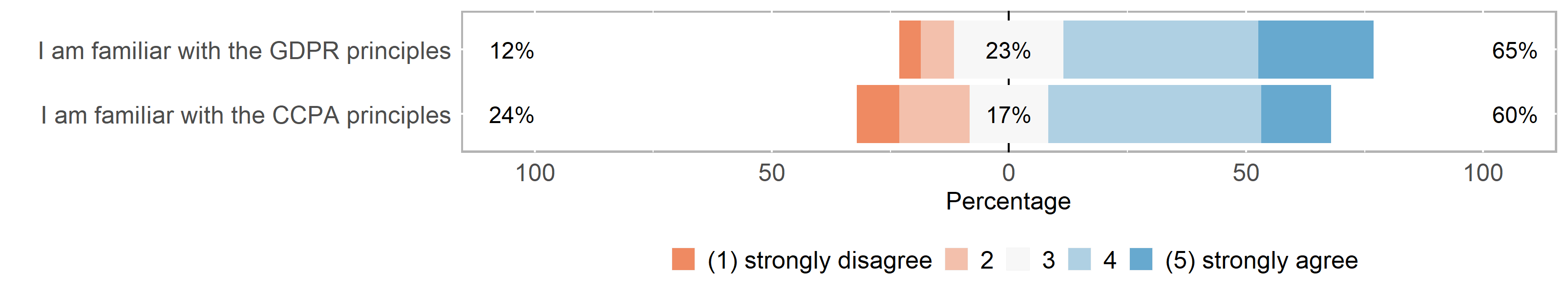}
\caption{Descriptive results of familiarity with privacy legislation.}
\label{fig:familiarity}
\end{figure}

\end{document}